# Measurement of the Fermi-LAT Localization Performance

**T.H. Burnett, M. Kerr, M. Roth**
on behalf of the Fermi/LAT collaboration
*University of Washington, Seattle WA 98195, USA*

We present results of a study of the localization capability of Fermi-LAT, using a large set of blazars with precise radio locations. Since the width of the PSF decreases with energy, the performance is typically dominated by a few high energy photons, so it is important to properly characterize the high-energy PSF. Using such data, we have found a need to modify the pre-launch high-energy (greater than a few GeV) PSF derived from extensive Monte Carlo simulations of particle interactions in the LAT; the resulting data-based PSF is shown.

## 1. INTRODUCTION

Associating sources seen by the LAT with other wavebands relies on size of the LAT "error box". Fig. 1, from, the very first LAT paper [1], shows how much better the LAT resolution is than EGRET.

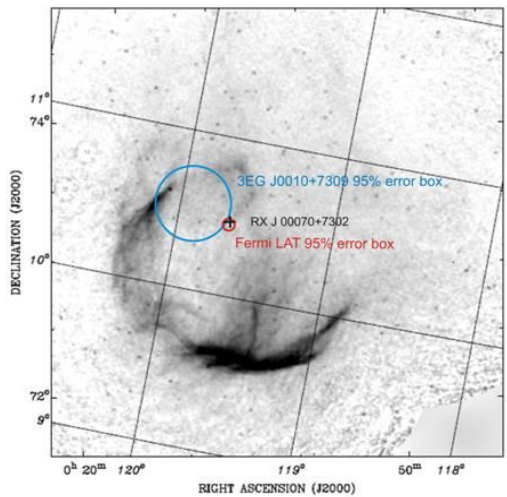

Figure 1: The *Fermi* LAT gamma-ray source, the central PWN X-ray source, and the corresponding *EGRET* source superimposed on a 1420 MHz map of CTA 1.

The data was taken during the first two months of operation, including post-launch checkout. Several factors contribute to this: the all-sky coverage and larger effective area provides more statistics on all sources, but the main improvement is the better high-energy performance of the LAT compared with previous detectors.

## 2. THE PSF

Fig. 2 illustrates the modular LAT design. Each of 16 modules has a tracker/converter above a calorimeter. An important consideration is that the tracker has two sections: 12 layers with 'thin' W foil converters above 4 layers with much thicker converters. Each has about the same effective area but the width of the Point Spread Function (PSF) in the thick section is 1.6x that of the thin section, both due to increased multiple scattering and due to the shorter length of tracks. Fig. 3 shows the nominal PSF used for these studies, determined using Monte Carlo modeling

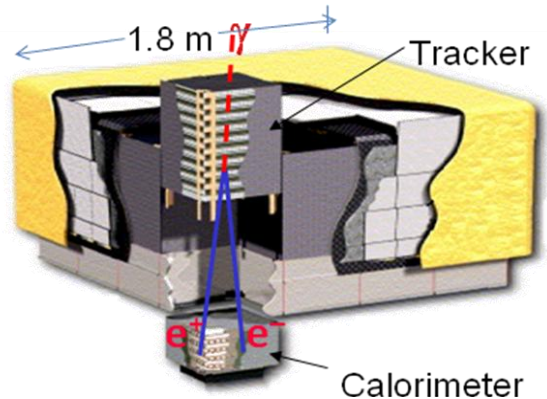

Figure 2: The LAT design, showing a blown-up module. The tracker actually has 18 layers, 12 "thin" front, 4 "thick" back, and 2 final layers without converters.

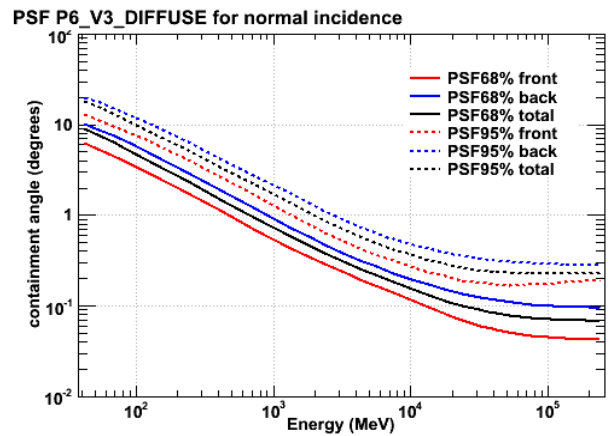

Figure 3: The LAT PSF, as measured using Monte Carlo simulation. This is for normal incidence, but the dependence on incidence angle, out to $60^\circ$, is minor.

Fig 3. illustrates two important features:

1. The resolution is a strong function of energy, due to multiple scattering of secondary $e^+e^-$ pair.

2. The front is better than the back

## 3. ANALYSIS USING BLAZARS

### 3.1. Initial selection

The analysis followed the steps:
- Select, as seed positions for further analysis, blazars from the BZCAT[2], CGRaBS[3], or CRATES[4] catalogs within 0.25 deg of any source in the preliminary 1 year catalog of LAT sources (640 found)
- Perform a region of Interest (ROI) analysis using 15 months of public data ("diffuse" class) for each source, assuming for a model:
  - a point source described by the P6_v3 PSF described above, and
  - background consisting of nearby catalog sources, the LAT standard galactic and isotropic diffuse distributions
- Define likelihood as a function of position, using four energy bands per decade from 1 GeV to 100 GeV, without assuming a spectral model. Maximize the likelihood for each energy band with respect to the signal.
- Fit the likelihood to a quadratic form. This procedure involves computation of a "quality" parameter, the goodness of the fit.
- Construct a "TS map" where 'TS' refers to the Test Statistic or twice the log likelihood, for a grid around the nominal position. The scale is defined by the fit. For evaluation, plot $\Delta TS$, with respect to the maximum, showing contours corresponding to 68%, 95%, 99% confidence region. (Colors chosen to show significance, fading to black at 5 σ.)

Fig 4. Shows an example TS map;

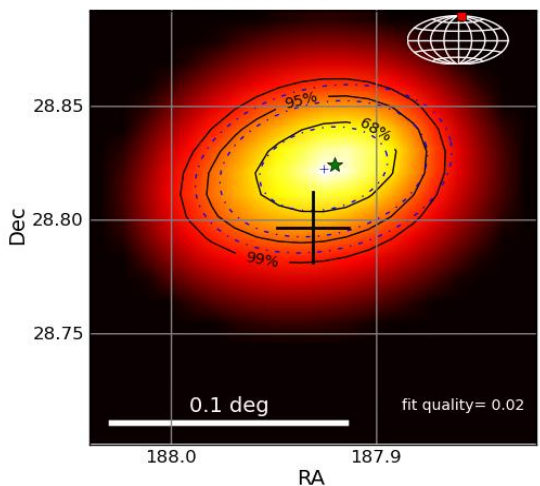

Figure 4: A "TS map", showing the point source significance as a function of position. Solid lines are the confidence region contours, dashed lines the equivalent contours from a quadratic fit to the surface. The star indicates the maximum $\Delta TS$, the large cross the seed position, corresponding to the blazar

### 3.2. Selection of Probable Associated Blazars

Apply the following requirements:
- High confidence ($\Delta TS>16$, probability of a random position exceeding this is less than 0.1%)(532 remain)
- Quadratic surface a good approximation (avoid multiple sources), quality parameter<1.0: (398 remain)
- Note that the position, other than the original 0.25 degree, is **not** a requirement.

Since the PSF is strongly energy-dependent, the "error box" depends on the spectral shape. The circular confidence radius plotted in Fig. 5 is the geometric mean of the elliptical axes. Note that the dependence on flux is consistent with an inverse square root

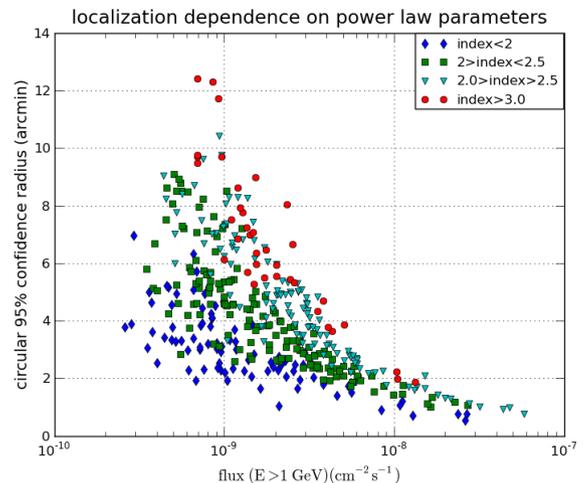

Figure 5: The dependence of the localization on the spectral index and flux.

### 3.3. Effective scale

Since the maximum likelihood is an estimator of the position, the shape is an *a posteriori* measure of the probability distribution for the actual position, given the data. This distribution is an exponential

$$\frac{dN}{d(\Delta TS)} \propto \exp\left(-\frac{\Delta TS}{2\,\zeta^2}\right)$$

where the factor $\zeta \approx 1$ is a scale multiplier, or "fudge" factor to account for the actual error (PSF) being larger by such a factor than predicted.

The value for $\zeta$ inferred from the fitting to the distribution is 1.10±0.05. Fig. 6 shows the distribution.

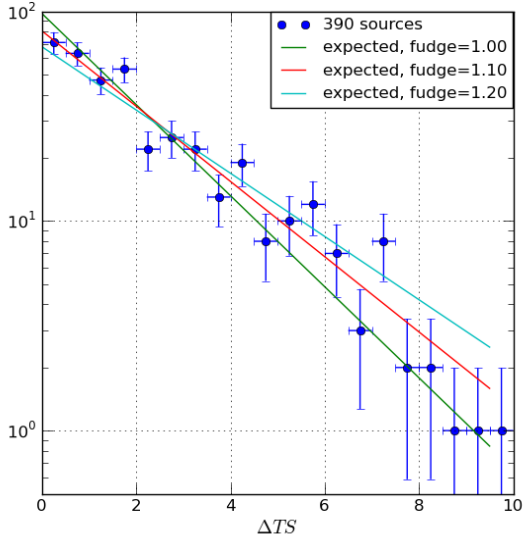

Figure 6: Distribution of the Test Statistic difference from the maximum to the AGN source. Lines showing the expected exponential distribution are shown.

## 4. A DIRECT MEASUREMENT OF THE PSF

The discrepancy implied by the need for the fudge factor can be verified by a direct measurement of the PSF. The following is a plot for E>32 GeV, using the same selected sources with $\Delta TS<9$. It shows that the PSF determined from the data is up to ~x2 wider than the Monte Carlo prediction at the highest energies for which we can measure it

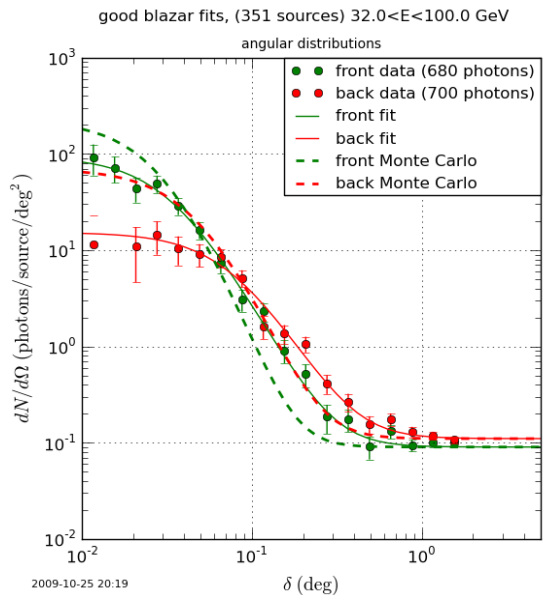

Figure 7: Example fit for deviations from blazer positions, for front and back photons

The fit function is

$$\frac{dN}{d\Omega}(\delta;\sigma,\gamma,s,b) = s\left(1-\frac{1}{\gamma}\right)\left(1+\frac{1}{2\gamma}\left(\frac{\delta}{\sigma}\right)^2\right)^{-\gamma} + b$$

where σ is a scale parameter, while γ describe the tail. *s* is number of signal counts, and *b is* the background density. The 68% containment depends on σ and γ, and is approximately 3 σ when γ=2.

The PSF is clearly not consistent with the Monte Carlo prediction. We have no explanation at this time, and will in the future use the measured value.

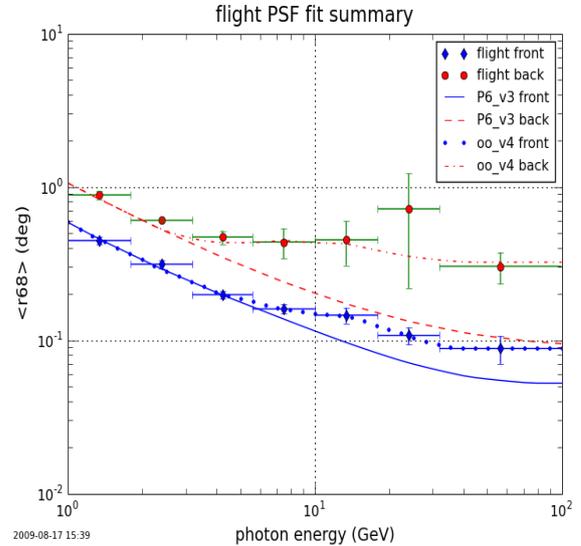

Figure 9: Summary of the 68% containment values implied by fits to the photon energy ranges shown.

Fig. 9 summarizes seven energy bands used for the fits, showing the deviation from the Monte Carlo prediction, starting around 4 GeV.

## 5. CONCLUSION

The need to modify the PSF, at least to optimize localization, is confirmed by measurements of the PSF, which departs from the Monte Carlo prediction above 4 GeV. An update for the PSF, based on data, is in preparation.

### Acknowledgments

We gratefully acknowledge support from NASA.
.

### References

[1] Abdo et al, *Science* 21 November 2008: Vol. 322. no. 5905, pp. 1218 – 1221
[2] Massaro et al. A&A **495 (2)** 691-696 (2009)
[3] Healey et al. 2008, ApJS, 175, 97
[4] Healey at al. The Astrophysical Journal Supplement Series, 171:61 Y 71, 2007 July